\documentclass[12pt]{article}
\begin{document}
\title{{\bf Cosmological Ontology and Epistemology}
\thanks{Alberta-Thy-20-14, arXiv:yymm.nnnn [hep-th], manuscript written for the proceedings of the Philosophy of Cosmology UK/US Conference, 12th\ --\ 16th September 2014, Tenerife, Spain}}

\author{Don N. Page
\thanks{Internet address:
profdonpage@gmail.com}
\\
Department of Physics\\
4-183 CCIS\\
University of Alberta\\
Edmonton, Alberta T6G 2E1\\
Canada}

\date{2014 December 23}
\maketitle
\large
\begin{abstract}
\baselineskip 25 pt

In cosmology, we would like to explain our observations and predict
future observations from theories of the entire universe.  Such
cosmological theories make ontological assumptions of what entities
exist and what their properties and relationships are.  One must also
make epistemological assumptions or metatheories of how one can test
cosmological theories.  Here I shall propose a Bayesian analysis in
which the likelihood of a complete theory is given by the normalized
measure it assigns to the observation used to test the theory.  In
this context, a discussion is given of the trade-off between prior
probabilities and likelihoods, of the measure problem of cosmology, of
the death of Born's rule, of the Boltzmann brain problem, of whether
there is a better principle for prior probabilities than mathematical
simplicity, and of an Optimal Argument for the Existence of God.

\end{abstract}
\normalsize
\baselineskip 20.7 pt
\newpage
\section{Introduction}

Cosmology is the study of the entire universe.  Ideally in science, one would like the simplest possible theory from which one could logically deduce a complete description of the universe.  Such theories must make implicit assumptions about the {\it ontology} of the universe, what are its existing entities and their nature.  They may also make implicit assumptions about the ontology of the entire world (all that exists), particularly if other entities beyond the universe have relationships with the universe.

For example, I am a Christian who believes that the universe was created by an omnipotent, omniscient, omnibenevolent, personal God who exists outside it but who relates to it as His creation.  Therefore, the ontology I assume for the world includes not only our universe but also God and other entities He may have created (such as new heavens and new earth for us after death).  However, my assumption that God has created our universe as an entity essentially separate from Himself means that one can also look for a theory of the universe itself, without including its relationship to God, though there may be some aspects of such a theory that can only be explained by the assumption that the universe was created by God.  (For example, even though a complete theory of our universe would necessarily imply the facts that there is life and consciousness within it, since such entities do exist within our universe, there may not be a good explanation of these aspects of the theory apart from the concept of creation by a personal God.)

Here I shall focus on theories of our universe as a separate entity, though near the end I shall speculate on some possible deeper explanations for why our universe is as it is.

I shall define a complete theory for a universe to be one that completely describes or specifies all properties of that universe.  I shall assume that an observer who could observe all properties of a universe and who has unlimited reasoning power could deduce a unique complete theory for that universe.  (Equivalent complete descriptions of the theory I regard as the same theory.  Depending upon one's background knowledge and the effort needed to deduce consequences from logically equivalent assumptions, different equivalent complete descriptions may be assigned different levels of simplicity when one seeks the simplest complete description, but that involves subjective elements that I shall not get into here.)

Of course, we do not know for certain what is the unique complete theory of our universe.  What we can know and how we can know it is the subject of {\it epistemology}.

Some of the complications of realistic epistemology arise from the limited reasoning powers of humans, such as the fact that we cannot deduce all the consequences of a set of axioms.  (For example, the axioms of arithmetic and of complex numbers presumably imply whether the Riemann hypothesis is true or false, but so far humans have not been able to find a rigorous deduction of which is the case.)  To avoid all the complications of limited reasoning power, for simplicity I shall just consider idealized cases in which observations are limited but reasoning powers are not.

Despite this idealization, we are also restricted by the fact that our observations are limited and do not show the entire universe.  A limited observation does not imply a unique complete theory for a universe.

Therefore, beings like us with limited observations within the universe cannot hope to attain absolute certainty about a complete theory for the universe.  The most we might hope for are posterior probabilities for different complete theories, probabilities taking into account our partial observations.  But even the idealized observer cannot deduce the posterior probabilities of different theories fitting his or her partial observations without making subjective choices for the prior probabilities of these different theories.

Here I shall lump all of the information accessible to the idealized observer when he or she assigns posterior probabilities to different theories into a single observation $O_k$, which includes memory elements of what in ordinary language one might consider to be many previous observations.  The observation $O_k$ represents all that the idealized observer knows about the universe before coming up with various complete theories $T_i$ to explain this observation and to which the idealized observer wishes to assign posterior probabilities $P(T_i|O_k)$ for the theory $T_i$ that are conditional upon the observation $O_k$.

\newpage

\baselineskip 18.7 pt

\section{A Bayesian Analysis for the Probabilities of Theories}

Consider theories $T_i$ that for each possible observation $O_k$ give a probability for that observation, $P(O_k|T_i)$.  This of course is generically {\it not} the same as the probability $P(T_i|O_k)$ of the theory given the observation, which is a goal of science to calculate.

If we assign (subjective) prior probabilities $P(T_i)$ to the theories $T_i$ (presumably higher for simpler theories, by Occam's razor) and use an observation $O_k$ to test the theory, $P(O_k|T_i)$ is then the likelihood of the theory, and by Bayes' theorem we can calculate the posterior probability of the theory as
\begin{equation}
P(T_i|O_k) = \frac{P(T_i)P(O_k|T_i)}{\sum_j P(T_j)P(O_k|T_j)}.
\label{Bayes}
\end{equation}
We would like to get this posterior probability as high as possible by choosing a simple theory (high prior probability $P(T_i)$) that gives a good statistical fit to one's observation $O_k$ (high likelihood $P(O_k|T_i)$).

Since a complete theory for a universe should completely specify all properties of that universe with certainty, one might think that it has no room for probabilities between 0 and 1, so that all likelihoods $P(O_k|T_i)$ would be 0 or 1.  Then the posterior probabilities of these theories, given the observation $O_k$, would just be proportional to the prior probabilities of the theories $T_i$ that give $P(O_k|T_i)=1$.

For observations to play a bigger role in science, we would like a way of getting a range of likelihoods, $0 < P(O_k|T_i) < 1$.  If the theories can give normalizable measures for the observations, then the normalized measures can be interpreted to be likelihoods $P(O_k|T_i)$ that can vary between 0 and 1.

In a classical theory, the measures could be functionals of the phase-space trajectory, such as the times that the system spends in different regions of the phase space.  In a quantum theory, the measures could be certain functionals of the quantum state (maps from the quantum state to nonnegative numbers).

Another argument against likelihoods all zero or one is the following:  If a specific theory $T_i$ leads to the definite existence of more than one observation $O_j$, one might be tempted to say that $P(O_j|T_i)=1$ for all of these observations, giving unit likelihood for any theory $T_i$ predicting the definite existence of the particular actual observation $O_k$ used to test the theory.  However, this procedure would mean that one could construct a theory of maximum (unit) likelihood, no matter what the observation turns out to be, just by having the theory predict that all possible observations definitely exist.  But this seems to be too cheap, supposedly explaining everything but actually explaining nothing.

Therefore, I propose the normalization principle or metatheory that each theory $T_i$ should give normalized probabilities for the different possible observations $O_j$ that it predicts, so that for each $T_i$, the sum of $P(O_j|T_i)$ over all $O_j$ is unity. 

One could view theory construction as a contest, in which theorists are given only unit probabilistic resources to divide at will for each theory, and then others judge the theories by assigning prior probabilities to the theories based on their simplicity.  Normalizing the products of these prior probabilities and of the probabilities the theories assign to the observation used to test the theory then gives the posterior probabilities of the theories.

For this to be a fair contest, each theory should be allowed a unit probability to distribute among all the observations predicted by the theory.  Of course, if the theorists cheat and give unnormalized probabilities in their theories (such as assigning probabilities near one for more than one observation in the same theory), the judges can compensate for that by assigning correspondingly lower prior probabilities to such theories.  However, it would give a more clear division between prior probabilities and likelihoods if theorists obeyed the proposed rules and normalized the observational probabilities for each theory so that they sum to one.

\section{Trade-Off Between Prior Probabilities and\\Likelihoods}

In a Bayesian analysis, we have (1) prior probabilities $P(T_i)$ for theories (intrinsic plausibilities), (2) conditional probabilities $P(O_k|T_i)$ for observations (`likelihoods' of the theories for a fixed observation), and (3) posterior probabilities $P(T_i|O_k) \propto P(T_i)P(O_k|T_i)$ for theories $T_i$, conditionalized upon the observation $O_k$.

Prior probabilities are subjective, usually higher for simpler theories.  The highest prior probability might be for the theory $T_1$ that nothing concrete (contingent) exists, but then $P(O_k|T_1)=0$.  $T_2$ might be the theory that all observations exist equally:  $P(O_k|T_2) = 1/\infty = 0$ when normalized (e.g., modal realism).

At the other extreme would be a maximal-likelihood theory giving $P(O_k|T_i)=1$ for one's observation $O_k$ (and zero probability for all other observations), but this seems to require a very complex theory $T_i$ that might be assigned an extremely tiny prior probability $P(T_i)$, hence giving a very low posterior $P(T_i|O_k)$.

Thus there is a trade off between prior probabilities and likelihoods, that is, between intrinsic plausibility and fit to observations.

\section{The Measure Problem of Cosmology}

The {\it measure problem} of cosmology (see, e.g., \cite{Linde:1993nz, Vilenkin:1994ua, Freivogel:2011eg, Page:2014eoa}) is how to obtain probabilities of observations from the quantum state of the universe.  This is particularly a problem when eternal inflation leads to a universe of unbounded size so that there are apparently infinitely many realizations or occurrences of observations of each of many different kinds or types, making the ratios ambiguous.  There is also the danger of domination by Boltzmann Brains, observers produced by thermal and/or vacuum fluctuations \cite{Dyson:2002pf, Albrecht:2002uz, Page:2006dt, Page:2006nt}.

The {\it measure problem} is related to the {\it measurement problem} of quantum theory, how to relate quantum reality to our observations that appear to be much more classical.  An approach I shall take is to assume that observations are fundamentally conscious perceptions or sentient experiences (each perception being all that one is consciously aware of at once).

I shall take an Everettian view that the wavefunction never collapses, so in the Heisenberg picture, there is one single fixed quantum state for the universe (which could be a `multiverse').  I assume that instead of `many worlds,' there are instead many different actually existing observations (sentient experiences) $O_k$, but that they have different positive measures, $\mu_k = \mu(O_k)$, which are in some sense how much the various observations occur, but they are {\it not} determined by just the contents of the observations.

For simplicity, I shall assume that there is a countable discrete set $\{O_k\}$ of observations, and that the total measure $\sum_k \mu_k$ is normalized to unity for each possible complete theory $T_i$ that gives the normalized measures $\mu_k$ for all possible observations $O_k$.  Then for a Bayesian analysis, I shall interpret the normalized measures $\mu_k$ of the observation $O_k$ that each theory $T_i$ gives as the probability of that observation given the theory, $P(O_k|T_i)$, which for one's observation $O_k$ is the `likelihood' of the theory $T_i$.

\section{Sensible Quantum Mechanics or Mindless Sensationalism}

The map from the quantum state to the measures of observations could be nonlinear.  However, I assume a linear relationship in Sensible Quantum Mechanics \cite{Page:1995dc, Page:1995nj, Page:1995kw, Page:1994pp, Page:2001ba, Page:2006er, Page:2011sr} (which I have also called Mindless Sensationalism because it proposes that what is fundamental is not minds but conscious perceptions, which crudely might be called `sensations,' though they include more of what one is consciously aware of, e.g., memories, than what is usually called `sensations'):
\begin{equation}
\mu(O_k) = \sigma[A(O_k)] \equiv \textrm{expectation value of the operator}\ A(O_k).
\label{observation-measure}
\end{equation}

Here $\sigma$ is the quantum state of the universe (a positive linear functional of quantum operators), and $A(O_k)$ is a nonnegative `awareness operator' corresponding to the observation or sentient experience (conscious perception) $O_k$.  The quantum state $\sigma$ (which could be a pure state, a mixed state given by a density matrix, or a C*-algebra state) and the awareness operators $\{A(O_k)\}$ (along with the linear relationship above and a description of the contents of each $O_k$) are all given by the theory $T_i$.

\section{The Death of the Born Rule}

Traditional quantum theory uses the Born rule with the probability of the observation $O_k$ being the expectation value of $A(O_k) = \mathbf{P}_k$ that is a projection operator ($\mathbf{P}_j\mathbf{P}_k = \delta_{jk}\mathbf{P}_k$, no sum over $k$) corresponding to the observation $O_k$, so 
\begin{equation}
P(O_k|T_i) = \sigma_i[\mathbf{P}_k] = \langle\mathbf{P}_k\rangle_i.
\label{Born}
\end{equation}

Born's rule works when one knows where the observer is within the quantum state (e.g., in the quantum state of a single laboratory rather than of the universe), so that one has definite orthonormal projection operators.  However, Born's rule does not work in a universe large enough that there may be identical copies of the observer at different locations, since then the observer does not know uniquely the location or what the projection operators are \cite{Page:2008ns, Page:2009qe, Page:2009mb, Page:2010bj}.

Why does the Born rule die?  Suppose there are two identical copies of the observer, at locations $B$ and $C$, that can each make the observations $O_1$ and $O_2$ (which do not reveal the location).  Born's rule would give the probabilities $P_1^B = \sigma[\mathbf{P}_1^B]$ and $P_2^B = \sigma[\mathbf{P}_2^B]$ if the observer knew that it were at location $B$ with the projection operators there being $\mathbf{P}_1^B$ and $\mathbf{P}_2^B$.  Similarly, it would give the probabilities $P_1^C = \sigma[\mathbf{P}_1^C]$ and $P_2^C = \sigma[\mathbf{P}_2^C]$ if the observer knew that it were at location $C$ with the projection operators there being $\mathbf{P}_1^C$ and $\mathbf{P}_2^C$.

However, if the observer is not certain to be at either $B$ or $C$, and if
$P_1^B < P_1^C$, then one should have $P_1^B < P_1 < P_1^C$.  But there is no state-independent projection operator that gives an expectation value with this property for all possible quantum states.  No matter what the orthonormal projection operators $\mathbf{P}_1$ and $\mathbf{P}_2$ are, there is an open set of states that gives expectation values that are not positively weighted means of the observational probabilities at the two locations.  Thus the Born rule fails in cosmology.

In more detail, consider normalized pure quantum states of the form
\begin{equation}
|\psi\rangle =  b_{12}|12\rangle + b_{21}|21\rangle, 
\label{2-state}
\end{equation}
with arbitrary normalized complex amplitudes $b_{12}$ and $b_{21}$.  The
component $|12\rangle$ represents the observation 1 in the region $B$ and the
observation 2 in the region $C$; the component $|21\rangle$ represents the
observation 2 in the region $B$ and 1 in the region $C$.  Therefore, $P_1^B =
P_2^C = |b_{12}|^2$, and $P_2^B = P_1^C = |b_{21}|^2$.

For Born's rule to give the possibility of both observational probabilities'
being nonzero in the two-dimensional quantum state space being considered, the
orthonormal projection operators must each be of rank one, of the form
\begin{eqnarray}
\mathbf{P}_1 = |\psi_{1}\rangle\langle\psi_{1}|,\ 
\mathbf{P}_2 = |\psi_{2}\rangle\langle\psi_{2}|,
\label{projections}
\end{eqnarray}
where $|\psi_{1}\rangle$ and $|\psi_{2}\rangle$ are two orthonormal pure states.

Once the state-independent projection operators are fixed, then if the
quantum state is $|\psi\rangle = |\psi_{1}\rangle$, the expectation values of
the two projection operators are $\langle\mathbf{P}_1\rangle \equiv
\langle\psi|\mathbf{P}_1|\psi\rangle =
\langle\psi_1|\psi_{1}\rangle\langle\psi_{1}|\psi_1\rangle = 1$ and
$\langle\mathbf{P}_2\rangle \equiv \langle\psi|\mathbf{P}_2|\psi\rangle =
\langle\psi_1|\psi_{2}\rangle\langle\psi_{2}|\psi_1\rangle = 0$.  These extreme
values of 1 and 0 are not positively weighted means of $P_1^B$ and $P_1^C$ and
of $P_2^C$ and $P_2^C$ for any choice of $|\psi_{1}\rangle$ and
$|\psi_{2}\rangle$ and any normalized choice of positive weights.  Therefore, no
matter what the orthonormal projection operators $\mathbf{P}_1$ and
$\mathbf{P}_2$ are, there is at least one quantum state (and actually an open
set of states) that gives expectation values that are not positively weighted
means of the observational probabilities at the two locations.  Thus Born's rule
fails.

The failure of the Born rule means that in a theory $T_i$, the awareness operators $A_i(O_k)$, whose expectation values in the quantum state $\sigma_i$ of the universe give the probabilities or normalized measures for the observations or sentient experiences $O_k$ as $P(O_k|T_i) \equiv \mu_i(O_k) = \sigma_i[A_i(O_k)] \equiv \langle A_i(O_k) \rangle_i$, cannot be projection operators.  However, the awareness operators could be weighted sums or integrals over spacetime of localized projection operators $\mathbf{P}_i(O_k,x)$ at locations denoted schematically by $x$, say onto brain states there that would produce the observations or sentient experiences.

\section{The Boltzmann Brain Problem}

In local quantum field theory with a definite globally hyperbolic spacetime, any positive localized operator (such as a localized projection operator) will have a strictly positive expectation value in any nonpathological quantum state. Therefore, if such a positive localized operator is integrated with uniform weight over a spacetime with infinite 4-volume, it will give an awareness operator with an infinite expectation value.

If one takes the integral only up to some finite cutoff time $t_c$ and normalizes the resulting awareness operators, then for a universe that continues forever with a 3-volume bounded below by a positive value, the integrals will be dominated by times of the same order of magnitude as the cutoff time.  If at late times the probability per 4-volume drops very low for ordinary observers, then most of the measure for observations will be contributed by thermal or vacuum fluctuations, so-called Boltzmann brains.  That is, Boltzmann brains will dominate the measure for observations.

If Boltzmann brains dominate the measure for observations, one might ask, ``So what?''  Couldn't it be that our observations are those of ordinary observers? Or couldn't it be that our observations really are those of Boltzmann brains?  However, since Boltzmann brain observations are produced mainly by thermal or vacuum fluctuations, it would be expected that only a very tiny fraction of their measure would be for observations so ordered as our observations.  This very tiny fraction, plus the even smaller fraction of ordered ordinary observer observations in comparison with the dominant disordered Boltzmann brain observations, would be only a very tiny fraction of the measure of all observations.  Thus the normalized probability of one of our ordered observations (which we would use as the likelihood of the theory) would be highly diluted and hence much smaller than those of alternative theories in which Boltzmann brains do not dominate.  If these alternative theories do not have prior probabilities that are too small, they would dominate the posterior probabilities.

In summary, Boltzmann brain domination, which is predicted by many simple extensions of current theories (e.g., with the awareness operators or their equivalent being obtained by a uniform integration over spacetime up to a cutoff that is then taken to infinity), gives a {\it reductio ad absurdum} for such theories, making their likelihoods very small.  Surely there are alternative theories that avoid Boltzmann brain domination without such a cost of complexity that their prior probabilities would be decreased so much as the gain in likelihoods from not having the normalized probabilities of our ordered observations highly diluted by disordered Boltzmann brain observations.

The Boltzmann brain problem is analogous to the ultraviolet catastrophe of late 19th century classical physics:  Physicists then did not believe that an ideal black body in thermal equilibrium would really emit infinite power, and physicists now do not believe that Boltzmann brains really dominate observations.

\section{Volume Weighting versus Volume Averaging}

The approach that gives `awareness operators' as uniform integrals over spacetime of localized projection operators (or equivalently counts all observation occurrences equally, not matter when and where they occur in a spacetime) gives an especially severe Boltzmann brain problem in spacetimes with a positive cosmological constant (as ours seems to have) with the spatial hypersurfaces having 3-volumes that asymptotically grow exponentially, as in the $k = 1$ slicing of the de Sitter spacetime.  At each time, counting the number or measure of observations as growing with the volume is called `volume weighting.'

In 2008 I proposed the alternative of Volume Averaging \cite{Page:2008zh, Page:2011gq}, which gives a contribution to the measure for an observation from a hypersurface that is proportional to the {\it spatial density} of the occurrences of the observation on the hypersurface.  This rewards the spatial frequency of observation occurrences rather than the total number that would diverge in eternal inflation as the hypersurface volume is taken to infinity.

Volume Averaging ameliorates the Boltzmann brain problem by not giving more weight to individual spatial hypersurfaces at very late times when Boltzmann brains might be expected to dominate.  However, when one integrates over a sequence of hypersurface with a measure uniform in the element of proper time $dt$, one gets a divergence if the time $t$ goes to infinity.  One needs some suppression at late times to avoid this divergence.

In 2010 I proposed Agnesi Weighting \cite{Page:2010re}, replacing $dt$ by $dt/(1+t^2)$ where $t$ is measured in Planck units.  This year I have also proposed new measures depending on the Spacetime Average Density (SAD) of observation occurrences within a proper time $t$ from a big bang or bounce \cite{Page:2014eoa}.  When these measures are combined with Volume Averaging and a suitable quantum state such as my Symmetric-Bounce one \cite{Page:2009ct}, they appear to be statistically consistent with all observations and seem to give much higher likelihoods than current measures using the extreme view that the measure is just given by the quantum state.  

\section{Is There a Better Principle than Mathematical Simplicity?}

Mathematical simplicity seems to be a reasonably good guide in science for choosing prior probabilities.  However, once observations are taken into account with the likelihoods, the highest posterior probabilities do not seem to go to the mathematically absolutely simplest theories (such as the theory that nothing concrete exists, which has zero likelihood given the fact that we observe something concrete, something not logically necessary).

Is there another principle that works better for predicting or explaining the properties of the actual world?

A conjectured principle for explaining the properties of the world is the following \cite{Page:2012wda}:

{\it The actual world is the best possible world.}

By the best possible world, I mean the one with maximum goodness.  I take the goodness that is maximized to be the pleasantness (measure of pleasure) of conscious or sentient experiences, which is what I am calling their intrinsic goodness.  Conscious experiences of pleasure (happiness, joy, contentment, satisfaction, etc.) would have positive goodness.  Conscious experiences of displeasure (unhappiness, pain, agony, discontentment, dissatisfaction, etc.) would have negative goodness.

Our universe seems to have enormous positive goodness, but it also seems to have enormous negative goodness.  Our universe also seems to  have a very high degree of mathematical elegance and beauty.  Humans can partially appreciate this, so that helps increase goodness.  But it would seem that intrinsic goodness consciously experienced by humans would be higher if disasters, disease, and cruelty were eliminated, even at the cost of less mathematical elegance and beauty for the laws of physics.

\section{The Optimal Argument for the Existence of God}

If the mathematical elegance of the universe were appreciated by a sentient Being outside the universe, that might increase the goodness of the world to a maximum, despite the sufferings within it.  Goodness might be maximized if the Being had all possible knowledge, leading to the hypothesis of omniscience for full appreciation.  Maximum goodness might also suggest the hypothesis that the Being has omnipotence for actualizing the best possible world.  If the Being actualizes a world of maximum goodness, one might postulate that the Being is a Creator and has omnibenevolence.  Such an omniscient, omnipotent, omnibenevolent Creator would fit the usual criteria for God.

Thus the assumption that the world has maximum goodness might suggest the conclusion of the existence of God:

Without God, it would seem that the goodness of the universe could be increased by violations of the laws of physics whenever such violations would lead to more pleasure within the universe.  However, with God, such violations might decrease God's pleasure so much that total goodness would be decreased.  Perhaps the actual world does maximize total goodness, despite suffering that is a consequence of elegant laws.

God may grieve over unpleasant consequences of elegant laws of physics and might even directly experience all of them Himself (as symbolized by the terrible suffering He experienced in the Crucifixion), but there may be that inevitable trade off that God takes into account in maximizing total goodness.  If God really does maximize total goodness, He is doing what is best.

Let me give a draft syllogism for one form of this Optimal Argument for the Existence of God:

{\it Assumptions}:

1.  The world is described by the simple hypothesis that it is the best, maximizing the pleasure within it.

2.  It is most plausible that either (a) our universe exists in isolation, or (b) our universe is created by God whose pleasure is affected by the universe and who has a nature determining what gives Him pleasure.

3.  Our universe could have had more pleasure.

4.  If God exists, it is possible that the total pleasure of the world (including both that within our universe and within God) is maximized subject to the constraint of His nature.

{\it Conclusions}:

5.  If our universe exists in isolation, 3 implies that it could have had more pleasure and hence the world could have been better, contradicting 1.

6.  Therefore, 1 and 3 imply that option (a) of 2 is false.

7.  Then 2 and 4 implies that it is most plausible that God exists and created our universe.

Of course, the assumptions of this argument are highly speculative, so it certainly does not give a proof of the existence of God from universally accepted axioms.  It is merely suggestive, hopefully motivating further investigations of other evidence, such as of historical records about the founders and key events in the development of the world religions.

Assumption 1 is motivated by Occam's razor but is modified slightly from the typical scientific form that one seeks theories that have the simplest mathematical formulation.  That form seems to work well when one constrains theories by observations, but it does not seem to give a fundamental explanation as to why our universe appears to be described by relatively simple mathematics but not the simplest possible mathematics.  In making the assumption of the best world, I assume that goodness is fundamentally given by the measure of pleasure (the pleasantness) of conscious experiences (with displeasure counting negatively).

Assumption 2 is highly debatable, since there are many other possibilities that people have considered.  Unless I made an expansion to include other possibilities, my argument might seem worthwhile mainly to those who are trying to choose between these two options.

Assumption 3 does seem rather obviously in agreement with our observations, at least if one considers all alternative logical possibilities for our universe.

Assumption 4 could be true in various different ways.  A traditional free-will defense of the problem of evil might assume that God gets sufficient pleasure from having persons in His universe with libertarian free will (e.g., so that they can love Him freely), so that such a world maximizes the total pleasure of God and of His creatures despite the sufferings (displeasures, which I am counting at negative pleasures) within the universe.  I am personally sceptical that it is logically possible for God to {\it create} totally from nothing creatures with libertarian free will, so I would instead postulate that God gets sufficient pleasure from a universe almost always obeying highly orderly and elegant laws of nature that He generally uses in His creation, so that the total pleasure of all conscience experiences (those of both God and His creatures) is maximized despite the sufferings that both God and His creatures also experience.  But one might also consider other alternative ways to flesh out how Assumption 4 could be true.

Once one makes these assumptions (and perhaps others that are implicit in my argument), then it does seem to me that the plausible existence of God as the Creator of our universe follows.  Of course, the existence of God does not depend on these assumptions; God could exist even if one or more of these assumptions are false, just as I personally believe in His existence while being highly sceptical of one or more of the assumptions of nearly all of the classical arguments for the existence of God.  But I do think that the Optimal Argument for the Existence of God gives a somewhat new slant on why it might be plausible to believe in God.  Furthermore, although one can do a Bayesian analysis for theories of our universe itself without reference to God, postulating the existence of God might help explain certain features of our best theories of our universe, such as why they are enormously more simple than they might have been but yet do not seem nearly so simple as what is purely logically possible.

\section{Conclusions}

I propose that in a Bayesian analysis in which the probability of a particular observation is used as the likelihood of the theory, the sum of the probabilities of all observations should be unity.

It seems plausible that one can find cosmological theories that avoid Boltzmann brain domination and explain the high order of our observations, though the {\it measure problem} is not yet solved.

The best theories do not seem to be the absolute simplest, so presumably something other than simplicity is maximized, such as goodness.  The best explanation for the actual world may be that it is the {\it best possible world}.  This assumption, with suitable assumptions about God's nature, such as love of mathematical elegance and of sentient beings, leads to the {\it Optimal Argument for the Existence of God}:

The world may maximize goodness only by including a God whose appreciation of the elegant universe and sentient beings He created overbalances the sufferings within the universe.

\section{Acknowledgments}

For my work on prior probabilities and likelihoods, on the measure problem of cosmology, on the death of Born's rule, and on the Boltzmann brain problem, I have benefited from discussions with many colleagues, including Andy Albrecht, Tom Banks, Raphael Bousso, Adam Brown, Steven Carlip, Sean Carroll, Brandon Carter, Willy Fischler, Ben Freivogel, Gary Gibbons, Steve Giddings, Daniel Harlow, Jim Hartle, Stephen Hawking, Simeon Hellerman, Thomas Hertog, Gary Horowitz, Ted Jacobson, Shamit Kachru, Matt Kleban, Stefan Leichenauer, Juan Maldacena, Don Marolf, Yasunori Nomura, Joe Polchinski, Steve Shenker, Eva Silverstein, Mark Srednicki, Rafael Sorkin, Douglas Stanford, Andy Strominger, Lenny Susskind, Bill Unruh, Erik Verlinde, Herman Verlinde, Aron Wall, Nick Warner, and Edward Witten.  Face-to-face conversations with Gary Gibbons, Jim Hartle, Stephen Hawking, Thomas Hertog, and others were enabled by the gracious hospitality of the Mitchell family and Texas A \& M University at a workshop at Great Brampton House, Herefordshire, England.  My scientific research has been supported in part by the Natural Sciences and Engineering Research Council of Canada.

For my work on the Optimal Argument for the Existence of God and related philosophical and theological ideas, I am indebted to conversations with David Albert, Michael Almeida, Luke Barnes, John Barrow, Nicholas Beale, Andrew Briggs, Peter Bussey, Bernard Carr, Sean Carroll, Brandon Carter, Khalil Chamcham, Robin Collins, Gary Colwell, William Lane Craig, Paul Davies, Stanley Deser, George Ellis, Peter Getzels, Shelly Goldstein, Stephen Hawking, Jim Holt, Colin Humphreys, John Leslie, Barry  Loewer, Klaas Kraay, Robert Lawrence Kuhn, Robert Mann, David Marshall, Thomas Nagel, Elliot Nelson, Timothy O'Conner, Cathy Page, Jason Pollack, Joel Primack, Carlo Rovelli, Simon Saunders, Michael Schrynemakers, Richard Swinburne, Donald Turner, Aron Wall, Christopher Weaver, David Wilkinson, William Wootters, Dean Zimmerman, Henrik Zinkernagel, and Anna Zytkow.  This does not by any means imply that all or any of these people agree with my ideas; I have perhaps benefited most from those who have expressed sharp disagreement.



\baselineskip 4pt

\begin{thebibliography}{99}

\bibitem{Linde:1993nz} 
  A.~D.~Linde and A.~Mezhlumian,
  ``Stationary Universe,''
  Phys.\ Lett.\ B {\bf 307}, 25 (1993)
  [gr-qc/9304015].

\bibitem{Vilenkin:1994ua} 
  A.~Vilenkin,
  ``Predictions from Quantum Cosmology,''
  Phys.\ Rev.\ Lett.\  {\bf 74}, 846 (1995)
  [gr-qc/9406010].

\bibitem{Freivogel:2011eg} 
  B.~Freivogel,
  ``Making Predictions in the Multiverse,''
  Class.\ Quant.\ Grav.\  {\bf 28}, 204007 (2011)
  [arXiv:1105.0244 [hep-th]].

\bibitem{Page:2014eoa} 
  D.~N.~Page,
  ``Spacetime Average Density (SAD) Cosmological Measures,''
  JCAP {\bf 1411}, 038 (2014),
  arXiv:1406.0504 [hep-th].

\bibitem{Dyson:2002pf} 
  L.~Dyson, M.~Kleban and L.~Susskind,
  ``Disturbing Implications of a Cosmological Constant,''
  JHEP {\bf 0210}, 011 (2002)
  [hep-th/0208013].

\bibitem{Albrecht:2002uz} 
  A.~Albrecht,
  ``Cosmic Inflation and the Arrow of Time,''
  in {\em Science and Ultimate Reality:  Quantum Theory, Cosmology, and Complexity}, edited by J.~D.~Barrow, P.~C.~W.~Davies, and
C.~L.~Harper, Jr.\ (Cambridge University Press, Cambridge, 2004), pp.\ 363-401,
  [astro-ph/0210527].

\bibitem{Page:2006dt} 
  D.~N.~Page,
  ``Is Our Universe Likely to Decay within 20 Billion Years?,''
  Phys.\ Rev.\ D {\bf 78}, 063535 (2008)
  [hep-th/0610079].
  
\bibitem{Page:2006nt} 
  D.~N.~Page,
  ``Is Our Universe Decaying at an Astronomical Rate?,''
  Phys.\ Lett.\ B {\bf 669}, 197 (2008)
  [hep-th/0612137].
  
\bibitem{Page:1995dc} 
  D.~N.~Page,
  ``Sensible Quantum Mechanics: Are Only Perceptions Probabilistic?''
  quant-ph/9506010.

\bibitem{Page:1995nj} 
  D.~N.~Page,
  ``Attaching Theories of Consciousness to Bohmian Quantum Mechanics,''
  quant-ph/9507006.

\bibitem{Page:1995kw} 
  D.~N.~Page,
  ``Sensible Quantum Mechanics: Are Probabilities Only in the Mind?''
  Int.\ J.\ Mod.\ Phys.\ D {\bf 5}, 583 (1996)
  [gr-qc/9507024].

\bibitem{Page:1994pp} 
  D.~N.~Page,
  ``Quantum Cosmology Lectures,''
  gr-qc/9507028.

\bibitem{Page:2001ba} 
  D.~N.~Page,
  ``Mindless Sensationalism: A Quantum Framework for Consciousness,''
  in {\em Consciousness: New Philosophical Perspectives}, 
  edited by Q.~Smith and A.~Jokic (Oxford, Oxford University Press, 2003), 
  pp.\ 468-506,
  quant-ph/0108039.

\bibitem{Page:2006er} 
  D.~N.~Page,
  ``Predictions and Tests of Multiverse Theories,''
  in {\em Universe or Multiverse?}, edited by
B.~J.~Carr (Cambridge University Press, Cambridge, 2007), pp.\ 411-429
  [hep-th/0610101].

\bibitem{Page:2011sr} 
  D.~N.~Page,
  ``Consciousness and the Quantum,''
  arXiv:1102.5339 [quant-ph].

\bibitem{Page:2008ns} 
  D.~N.~Page,
  ``Insufficiency of the Quantum State for Deducing Observational Probabilities,''
  Phys.\ Lett.\ B {\bf 678}, 41 (2009)
  [arXiv:0808.0722 [hep-th]].

\bibitem{Page:2009qe} 
  D.~N.~Page,
  ``The Born Rule Dies,''
  JCAP {\bf 0907}, 008 (2009)
  [arXiv:0903.4888 [hep-th]].

\bibitem{Page:2009mb} 
  D.~N.~Page,
  ``Born Again,''
  arXiv:0907.4152 [hep-th].

\bibitem{Page:2010bj} 
  D.~N.~Page,
  ``Born's Rule Is Insufficient in a Large Universe,''
  arXiv:1003.2419 [hep-th].

\bibitem{Page:2008zh} 
  D.~N.~Page,
  ``Cosmological Measures without Volume Weighting,''
  JCAP {\bf 0810}, 025 (2008)
  [arXiv:0808.0351 [hep-th]].
  
\bibitem{Page:2011gq} 
  D.~N.~Page,
  ``Cosmological Measures with Volume Averaging,''
  Int.\ J.\ Mod.\ Phys.\ Conf.\ Ser.\  {\bf 01}, 80 (2011).

\bibitem{Page:2010re} 
  D.~N.~Page,
  ``Agnesi Weighting for the Measure Problem of Cosmology,''
  JCAP {\bf 1103}, 031 (2011)
  [arXiv:1011.4932 [hep-th]].

\bibitem{Page:2009ct} 
  D.~N.~Page,
  ``Symmetric-Bounce Quantum State of the Universe,''
  JCAP {\bf 0909}, 026 (2009)
  [arXiv:0907.1893 [hep-th]].

\bibitem{Page:2012wda} 
  D.~N.~Page,
  ``The Everett Multiverse and God,''
  in K.~Kraay, ed.\ ``God and the Multiverse: Scientific, Philosophical, and Theological Perspectives,'' Routledge Studies in the Philosophy of Religion (Book 10), London: Routledge (2014),
  arXiv:1212.5608 [physics.gen-ph].

\end{thebibliography}
\end{document}